\documentstyle[]{article}
\font\cero=cmss10 scaled 1728
\font\uno=cmssbx10 scaled 1200
\begin{document}
\begin{flushleft}
{\cero The Euler characteristic and the first Chern number in the covariant
phase space formulation of string theory} \\[3em]
\end{flushleft}
{\sf R. Cartas-Fuentevilla}\\
{\it Instituto de F\'{\i}sica, Universidad Aut\'onoma de Puebla,
Apartado postal J-48 72570, Puebla Pue., M\'exico (rcartas@sirio.ifuap.buap.mx).}  \\[4em]

Using a covariant description of the geometry of deformations for
extendons, it is shown that the topological corrections for the
string action associated with the Euler characteristic and the
first Chern number of the normal bundle of the worldsheet,
although do not give dynamics to the string, modify the symplectic
properties of the covariant phase space of the theory. Future
extensions
of the present results are outlined. \\


\noindent {\uno I. Introduction} \vspace{1em}

As it is well known, string theory contains two natural
topological invariants related with the different topologies of
the two-dimensional world surface embedded in a background
spacetime. Specifically, the Gauss-Bonnet action, which depends
only on the purely intrinsic properties of the world surface,
corresponds to the Euler characteristic, which counts the number
of holes or handles of the world surface. Additionally, the first
Chern number of the normal bundle of the world surface, which
depends on the extrinsic properties of the world surface embedded
in a (four-dimensional) background spacetime, gives us essentially
the number of self-intersections of the world surface. Such
topological invariants do not contribute effectively as Lagrangian
terms to the string dynamics, although it is well known also that
there is a global contribution in the path integral formulation of
the theory, weighting the different topologies in the sum over
world surfaces.

On the other hand, in the canonical formulations of the theory for
quantization, based on the classical dynamics of the theory, the
topological terms will have no any contribution, since the
dynamics remains unmodified. The fact that such terms play a
nontrivial role in the path integral formulation of the theory,
and do not appear at all in the canonical scheme, is somewhat
suspicious at first glance. Hence, the aim of this work is to show
that, on the basis of a covariant description of the canonical
formulation of the theory, the topological terms in the string
action may have indeed a physically relevant contribution on the
symplectic structure constructed on the corresponding covariant
phase space. With these results, we give a new relevant role of
the topological terms within a canonical formulation, which is
completely unknown in the literature, levelling thus the roles of
such terms in both approaches for quantization.

In the next section, we outline the covariant canonical formalism,
and in Sec. III we give some remarks on the covariant canonical
formulation of the Dirac-Nambu-Goto (DNG) action, in order to
prepare the background for the subsequent sections, where the
topological terms will be worked out. In Sec. VII we conclude
with some discussions about our results and future extensions. \\[2em]

\noindent {\uno II. Covariant phase space and the exterior
calculus} \vspace{1em}

In this section we summarize the exterior calculus on the
covariant phase space given in Ref. \cite{1} but adjusted for the
treatment of embeddings \cite{2}.

According to Ref. \cite{1}, in a given physical theory, {\it the
classical phase space is the space of solutions of the classical
equations of motion}, which corresponds to a manifestly covariant
definition. The basic idea of the covariant description of the
canonical formalism is to construct a symplectic structure on such
a phase space, instead of choosing {\it p's} and {\it q's}.

In the present case, the phase space is the space of solutions of
 (12), and we shall call it $Z$. Any background quantity, will be
associated with zero-forms on $Z$. The deformation operator
$\delta$ (see Sec. III) acts as an exterior derivative on $Z$,
taking $k$-forms into $(k+1)$-forms, and it should satisfy
\begin{equation}
     \delta^{2} = 0,
\end{equation}
and the Leibniz rule
\begin{equation}
     \delta (AB) = \delta AB + (-1)^{A} A \delta B.
\end{equation}
In particular, $\delta X^{\mu}$ is the exterior derivative of the
zero-form $X^{\mu}$ [see Eq.\ (5)], and it will be closed,
\begin{equation}
     \delta^{2} X^{\mu} = 0.
\end{equation}
Furthermore, since $\phi^{i} = n^{i}_{\mu} \delta X^{\mu}$, and
$n^{i}_{\mu}$ corresponds to  zero-forms on $Z$, the scalar fields
$\phi^{i}$ are one-forms on $Z$, and thus are anticommutating
objects: $\phi^{i}\phi^{j} = - \phi^{j}\phi^{i}$. This property
allows us to verify that, being the vector field $\delta = n^{i}
\phi_{i}$, thus $\delta^{2} = n^{i} n^{j} \phi_{i}\phi_{j}$, which
vanishes because of the commutativity of the zero-forms $n^{i}$
and the anticommutativity of the $\phi^{i}$ on $Z$, in full
agreement with Eq.\ (1). It is important to mention, at this
point, that the covariant deformation operator $D_{\delta}$ (see
Sec. III) also works as an exterior derivative on $Z$, in the
sense that maps $k$-forms into $(k+1)$-forms; however
$D^{2}_{\delta}$ does not vanish necessarily.

We can determine certain two-forms on $Z$ that will be useful for
our present purposes. Considering that $\delta \equiv \delta
X^{\mu} \partial_{\mu}$ and $D_{\delta} \equiv \delta X^{\mu}
D_{\mu}$, we can show that $D_{\delta} (\delta X^{\mu})$ vanishes
\begin{equation}
     D_{\delta} (\delta X^{\mu}) = \delta X^{\alpha} D_{\alpha}
\delta X^{\mu} = \delta X^{\alpha} [ \partial_{\alpha} \delta
X^{\mu} + \Gamma^{\mu}_{\alpha\lambda} \delta X^{\lambda}] =
\delta^{2} X^{\mu} + \Gamma^{\mu}_{\alpha\lambda} \delta
X^{\alpha} \delta X^{\lambda} = 0,
\end{equation}
where the first term vanishes according to Eq.\ (3), and the
second one because of the symmetry of
$\Gamma^{\mu}_{\alpha\lambda}$ in the indices $\alpha$ and
$\lambda$, and the anticommutativity of $\delta X^{\alpha}$ and
$\delta X^{\lambda}$. Hence, Eq.\ (4) suggests that $D_{\delta}$
is, as well as $\delta$, a measure of the closeness of $\delta
X^{\mu}$ on $Z$.  \\[2em]

\noindent {\uno III. Covariant canonical formulation for DNG {\it
p}-branes in a curved background} \vspace{1em}

It will be convenient to do the general treatment for {\it
p}-branes, and then to consider the particular case of string
theory (1-brane), which will show the particularities of string
theory as opposed to the other higher-dimensional objects.

In Ref. \cite{2}, it is shown that there exists an identically
closed two-form on the space of solutions of the classical
equations of motion (modulo gauge transformations) for {\it
p}-branes propagating in a curved background, endowing to the
physical phase space $\hat{Z}$ of a symplectic structure. However,
a more detailed study \cite{3} shows that such a closed two-form
is even an exact two-form, obtained by direct exterior derivative
of a {\it symplectic potential}, a global one-form on the phase
space. The strategy in Ref. \cite{3} for obtaining the global
symplectic potential directly from the variations of the
corresponding Lagrangian is as follows.

In the geometry of deformations of {\it p}-branes developed in
Ref. \cite{4}, it is assumed that an infinitesimal deformation
tangent to the world surface is not physically relevant, since it
can be identified always with the action of a world surface
diffeomorphism. However, as claimed in Ref. \cite{3}, it is
precisely such an infinitesimal diffeomorphism that plays the role
of our global symplectic potential on the phase space (and it is
the first example showing that a {\it spurious} quantity in a
conventional sense, may be physically relevant on the phase
space). Hence, we will maintain explicitly a world surface
diffeomorphism from the beginning, modifying slightly the original
deformation scheme given in Ref. \cite{4}.

In this manner, following Ref. \cite{4}, the deformation of the
world surface swept out by a {\it p}-brane (propagating in a
curved background) is given by the infinitesimal spacetime
variation
\begin{equation}
    \xi^{\mu} \equiv \delta X^{\mu} = n^{\mu}_{i} \phi^{i} +
    e^{\mu}_{a} \phi^{a},
\end{equation}
where $n_{i}$ correspond to vector fields normal to the world
surface and, $e_{a}$ to the vector fields tangent to such a
surface \cite{4}. Hence, considering that $D_{\mu}$ is the
background torsionless covariant derivative, in Ref. \cite{4} the
normal deformation operator is defined as
\begin{equation}
    D_{\delta} \equiv \delta^{\mu} D_{\mu}, \qquad \delta \equiv n_{i}
    \phi^{i},
\end{equation}
and it is found that \cite{4}
\begin{eqnarray}
    D_{\delta} e_{a} \!\! & = & \!\! (K_{ab}{^{i}} \phi_{i}) e^{b} +
    (\widetilde{\nabla}_{a} \phi_{i}) n^{i}, \nonumber\\
    D_{\delta} \gamma_{ab} \!\! & = & \!\! 2 K_{ab}{^{i}} \phi_{i},
    \nonumber \\
    D_{\delta} \sqrt{-\gamma} \!\! & = & \!\! \sqrt{-\gamma} K^{i}
    \phi_{i},
\end{eqnarray}
which will be useful below. We define here the tangential
deformation operator as
\begin{equation}
    D_{\Delta} \equiv \Delta^{\mu} D_{\mu}, \qquad \Delta \equiv
    e_{a} \phi^{a},
\end{equation}
and using the generalized Gauss-Weingarten equations, we find that
\begin{eqnarray}
    D_{\Delta} e_{a} \!\! & = & \!\! (\nabla_{a} \phi^{b}) e_{b} -
    K_{ab}{^{i}}\phi^{b} n_{i}, \nonumber\\
    D_{\Delta} \gamma_{ab} \!\! & = & \!\! \nabla_{a} \phi_{b} +
    \nabla_{b} \phi_{a}, \nonumber\\
    D_{\Delta} \sqrt{-\gamma} \!\! & = & \!\! \sqrt{-\gamma} \nabla_{a}
    \phi^{a}.
\end{eqnarray}
In this manner, the action for DNG {\it p}-branes,
\begin{equation}
    S_{0} = - \sigma_{0} \int d^{D} \xi \sqrt{-\gamma},
\end{equation}
considering world surface diffeomorphisms, has as first variation
\begin{equation}
    0 = (D_{\delta} + D_{\Delta}) S_{0} = - \sigma_{0} \int d^{D}
    \xi \sqrt{-\gamma} K^{i} \phi_{i} - \sigma_{0} \int d^{D}
    \xi \sqrt{-\gamma} \nabla_{a} \phi^{a},
\end{equation}
where the last of Eqs.\ (7) and (9) have been considered; from
Eq.\ (11) we can see that $D_{\Delta} S_{0}$ is associated with a
total divergence that can be indeed negligible, since it does not
contribute locally to the dynamics.

Therefore, from Eq.\ (11), the equations of motion for extremal
surfaces are
\begin{equation}
    K^{i} = 0,
\end{equation}
whose set of solutions defines, in fact, the covariant phase space
$Z$ of the theory.

On the other hand, from the deformation dynamics [obtained
linearizing Eq.\ (12)], and using the scheme of (self-)adjoint
operators, it is found that the current $j^{a} = \phi_{i}
\widetilde{\nabla}^{a} \phi^{i}$ is world surface covariantly
conserved $(\nabla_{a} j^{a} = 0)$, and corresponds to a closed
two-form on the phase space $D_{\delta} j^{a} = 0$ \cite{2}.
Therefore, one can finally construct a covariant and gauge
invariant symplectic structure $\omega$ for the theory \cite{2},
\begin{equation}
    \omega \equiv \int_{\Sigma} \sqrt{-\gamma} j^{a} d \Sigma_{a},
\end{equation}
independent on the choice of $\Sigma$ (a spacelike section of the
world surface corresponding to a Cauchy {\it p}-surface for the
configuration of the {\it p}-brane). However, the symplectic
current $j^{a}$ is even an exact two-form \cite{3}, since from
Eqs.\ (4), (5), and the first of Eqs.\ (7) one finds that
\begin{equation}
    \delta \phi^{a} = D_{\delta} (e^{a}_{\mu} \delta X^{\mu}) = -
    j^{a},
\end{equation}
and $j^{a}$ is in particular a closed form, $\delta j^{a} = 0$,
because of the nilpotency of $\delta$ [see Eq.\ (1)]. Therefore
$\phi^{a}$ (the tangential projection of the deformation of the
embedding), coming directly from the pure divergence term in Eq.\
(11), plays the role of a global symplectic potential on the phase
space. As pointed out above, $\phi^{a}$ gives no dynamics to the
string, but it is physically relevant on the phase space, in
accordance with Eq.\ (14).

It is important to emphasize here, the significance of a covariant
and gauge invariant symplectic structure for the theory. $\omega$
in Eq.\ (13) represents a complete {\it Hamiltonian} description
of the covariant phase space, preserving manifestly all relevant
symmetries of the theory. Hence, $\omega$ represents a starting
point for the study of the symmetry aspects and also a covariant
description of the canonical formulation of the theory for
quantization. For the general features of the covariant phase
space formulation, see for example Ref. \cite{1}. Note, however,
that the concept of symplectic potential does not appear at all in
Ref. \cite{1}.

In the next section we will follow the same procedure employed in
this section for determining the symplectic potential $\phi^{a}$,
in order to find the contribution of the Gauss-Bonnet topological
term in the action on the covariant phase space of the theory. \\[2em]

\noindent {\uno IV. The Gauss-Bonnet topological term}
\vspace{1em}

The Gauss-Bonnet term for an arbitrary closed {\it p}-brane
without physical boundaries is proportional to the Ricci scalar
$\cal R$ constructed from the world surface metric $\gamma_{ab}$,
\begin{equation}
    \chi \equiv \sigma_{1} \int d^{D} \xi \sqrt{-\gamma} {\cal R},
\end{equation}
whose first variation, according to the geometry of deformations
of Ref.\cite{4}, is given by
\begin{equation}
    D_{\delta} (\sqrt{-\gamma} {\cal R}) = - 2 \sqrt{-\gamma} G_{ab}
    K^{ab}{_{i}} \phi^{i} + \sqrt{-\gamma} \nabla_{a} \psi^{a}_{N},
\end{equation}
where $G_{ab}$ is the world surface Einstein tensor
\begin{equation}
    G_{ab} = {\cal R}_{ab} -\frac{1}{2} \gamma_{ab} {\cal R},
\end{equation}
and
\begin{equation}
    \psi^{a}_{N} = \gamma^{ab} D_{\delta} \gamma_{bc}^{c} - \gamma^{bc}
    D_{\delta} \gamma_{bc}^{a},
\end{equation}
is the analogous to $\phi^{a}$ in Section III. Note that in Ref.
\cite{4}, the pure divergence term involving $\psi^{a}_{N}$ is
completely ignored (which is correct in a conventional analysis of
the brane dynamics), without suspecting the relevant role that
such a term will play in a phase space formulation of the theory.

In this manner, Eq.\ (16) gives the universal contribution of the
Gauss-Bonnet term on the brane dynamics (the first term on the
right-hand side), and on the covariant phase space through
$\psi^{a}_{N}$. In this sense, there is no surprise for an
arbitrary {\it p}-brane, since $\chi$ changes the dynamics and
correspondingly the symplectic structure on the phase space.
Nevertheless, as it is well known, in a two-dimensional world
surface (and only for such a case), swept out for a string, the
world surface Einstein tensor vanishes (for an arbitrary embedding
background dimension),
\begin{equation}
    G_{ab} = 0,
\end{equation}
and there is not effect on the string dynamics. However, there is
a nontrivial contribution of the topological term on the phase
space through symplectic potential (18), independently on the null
contribution at the level of the string dynamics. For example, in
the case considered in Sec. III, if Eqs.\ (12) are the equations
of motion for the closed string dynamics described by external
surfaces, the inclusion of the topological term $\chi$ in the
action leaves Eqs.\ (12) unchanged (and thus the phase space
itself, defined as the space of solutions of the equations of
motion, is unaltered), but the corresponding symplectic potential
on the phase space is no longer $-\sigma_{0} \phi^{a}$, but
$-\sigma_{0} \phi^{a} + \sigma_{1} \psi^{a}_{N}$, where
$\psi^{a}_{N}$ is given in Eq.\ (18).

Although we have considered only the DNG closed strings as the
reference Lagrangian term for the inclusion of the GB term, the
main idea is to show that $\sigma_{1} \psi^{a}_{N}$ is the
universal contribution of the latter  on the phase space, and
similarly for any action describing strings, for example, some
action including terms with curvature corrections \cite{3}.
Therefore, if $\Phi^{a}$ is the symplectic potential for such a
general action, we can construct the symplectic structure $\omega$
as
\begin{equation}
    \omega = \int_{\Sigma} D_{\delta} \sqrt{-\gamma} (\Phi^{a} +
    \sigma_{1} \psi^{a}_{N}) d \Sigma_{a},
\end{equation}
with the wanted properties of closeness ($\delta\omega = 0$). The
closeness of $\omega$ is equivalent to the Jacobi identity that
the Poisson brackets satisfy, in the usual Hamiltonian scheme
\cite{5}.

Considering the deformation formulas of Ref. \cite{4}, we can
determine explicitly the universal contribution of $\psi^{a}_{N}$
to the symplectic current of the theory, in terms of $\phi^{i}$,
the only measure of the deformation that cannot be gauged away,
\begin{equation}
    D_{\delta} \psi^{a}_{N} = -2 \phi_{i} \{ K^{abi} \nabla_{b} (K^{j}
    \phi_{j}) + K^{bci} [\nabla_{b} (K_{c}^{aj} \phi_{j}) + \nabla_{c}
    (K_{b}^{aj} \phi_{j}) - \nabla^{a} (K_{bc}^{j} \phi_{j})]\};
\end{equation}
which will constitute the universal integral kernel of the Euler
characteristic on the symplectic geometry of string theory.
 Note that even in the simplest case of a DNG closed string
dynamics described by Eq.\ (12), $D_{\delta} \psi^{a}_{N}$ in Eq.\
(21) does not vanish. Therefore, the topological term modifies
drastically the symplectic properties of the phase space of the
theory, without changing the dynamics and the phase space itself.
\\ [2em]

\noindent {\uno V. The first Chern number of the normal bundle of
the worldsheet} \vspace{1em}

The self-intersection number of the worldsheet embedded in a
four-dimensional background spacetime, given essentially by the
first Chen number of the normal bundle of the worldsheet, has the
analytic expression \cite{6,7}
\begin{equation}
     \nu = \sigma_{2} \int d^{2} \xi \Omega,
\end{equation}
in terms of the extrinsic twist curvature:
\begin{eqnarray}
     \Omega \!\! & = & \!\! \frac{1}{2} \epsilon_{ij} \epsilon^{ab}
     \Omega_{ab}{^{ij}}, \\
     \Omega_{ab}{^{ij}} \!\! & = & \!\! \partial_{b}
     \omega_{a}{^{ij}} - \partial_{a} \omega_{b}{^{ij}}, \nonumber
\end{eqnarray}
where $\omega_{a}{^{ij}}$ corresponds to the extrinsic twist
potential \cite{4,7}. Using Eqs.\ (23), $\nu$ can be rewritten
explicitly as a topological invariant in terms of a total
divergence,
\begin{equation}
     \nu = \sigma_{2} \int d^{2} \xi \sqrt{-\gamma} \nabla_{a}
     (\epsilon_{ij} \varepsilon^{ab} \omega_{b}{^{ij}}),
\end{equation}
where $\epsilon^{ab} = \sqrt{-\gamma} \varepsilon^{ab}$ \cite{7}.
In order to determine the variation of $\nu$, we exploit the frame
gauge dependence of the potential $\omega_{b}{^{ij}}$, which means
that it can always be set equal to zero at any single chosen point
by an appropriate choice of the normal frame,
\begin{equation}
     \delta \nu = \sigma_{2} \int d^{2} \xi \sqrt{-\gamma} \nabla_{a}
     (\epsilon_{ij} \varepsilon^{ab} \delta \omega_{b}{^{ij}}),
\end{equation}
where the (normal) variation of the twist potential is given in
terms of $\phi^{i}$ by \cite{4,7}
\begin{equation}
     \delta \omega_{b}{^{ij}} = -2 K_{cb}{^{i}}
              \widetilde{\nabla}^{c} \phi^{j} + R_{\mu\nu\alpha\beta} n^{\mu
     i} n^{\nu j} n^{\alpha\kappa} e^{\beta}_{b} \phi_{\kappa},
\end{equation}
where $R_{\mu\nu\alpha\beta}$ is the Riemann tensor of the
(four-dimensional) background spacetime. In this manner, following
the ideas of the present work, from Eq.\ (25) we can identify to
$\Theta^{a} = \sigma_{2} \epsilon_{ij} \varepsilon^{ab} \delta
\omega_{b}{^{ij}}$ as a symplectic potential for $\nu$. Thus, the
contribution of $\nu$ on the integral Kernel of the symplectic
structure of the theory is given by the deformation (exterior
derivative) of $\Theta^{a}$,
\begin{equation}
    \delta \Theta^{a} = - K_{i} \phi^{i} \Theta^{a},
\end{equation}
where we have considered that $\delta\sqrt{-\gamma} =
\sqrt{-\gamma} K_{i} \phi^{i}$ \cite{4}. Note that the effect of
adding $\nu$ to the DNG string action is, in addition to leave
unaltered the dynamics governed by $K^{i}=0$ [Eq.\ (12)], to leave
unchanged the symplectic structure for the theory (unlike the case
of $\chi$ in Sec. IV), since in this case $\delta\Theta^{a} = 0$
in accordance with Eq.\ (27); of course the situation is different
in a more general case than that described by the DNG action,
where in general $K^{i} \neq 0$. Therefore, if $\Psi^{a}$ is the
symplectic potential for such a general action (which may include,
for example, the Gauss-Bonnet term $\chi$ considered in Sec. IV),
the symplectic structure $\omega$ of string theory (in four b
dimensions) including the term $\nu$ will take the form
\begin{equation}
    \omega = \int_{\Sigma} D_{\delta} (\sqrt{-\gamma} \Psi^{a})
    d\Sigma_{a} + \int_{\Sigma} D_{\delta} \Theta^{a} d\Sigma_{a},
\end{equation}
which is evidently closed.

The same argument employed in Ref. \cite{2} for demonstrating the
nondegeneracy of the symplectic structure for DNG branes works for
the contributions of the topological terms on the symplectic
structure of the theory. In this manner the $\omega$'s in (20) and
(28) are nondegenerate  and are defined on the reduced phase space
$Z/G$, with $G$ being the volume of the group of infinitesimal
spacetime diffeomorphisms \cite{2}. \\[2em]

\noindent {\uno 6. Remarks on open and closed strings}
\vspace{1em}

In Ref. \cite{7} open strings with topologically inspired boundary
conditions are considered, specifically the topological terms
considered here for closed strings. In both cases, such
topological terms do not affect the equations of motion; however,
in the case of open strings such terms lead to boundary conditions
to be implemented in addition to the equations of motion, unlike
the case treated here of closed strings, where such complementary
conditions do not appear at all, because of the absence of
physical boundaries. Thus, it is opportune to emphasize the
results presented in this work: the topological terms for closed
strings do not modify the classical dynamics, neither imposing any
additional condition, but contributing explicitly to the
symplectic structure of the theory, and thus they may have
possible quantum effects.

It is important to remark also that the symplectic potentials for
the topological terms $\chi$ and $\nu$ are the same for both
closed and open strings; however for the former the symplectic
structure will be constructed on the phase space defined by the
(unmodified) equations of motion, and for the latter on a {\it
restriction} of the same phase space defined by the complementary
conditions mentioned above. \\[2em]

\noindent {\uno 7. Remarks and prospects} \vspace{1em}

In Ref. \cite{8}, it is proved that the topological terms modify
drastically the deformation dynamics of string theory, in such a
way that the symplectic current obtained as a consequence of the
self-adjointness of that deformation dynamics, is in full
agreement with the currents obtained in the present treatment
calculating the variations of the symplectic potentials.
Specifically it is proved that the symplectic currents of the
topological terms in Eqs.\ (20) and (28) are worldsheet
covariantly conserved: $\nabla_{a} (D_{\delta}\sqrt{-\gamma}
\psi^{a}_{N}) = 0 = \nabla_{a} (D_{\delta} \Theta^{a})$, which
ensures that the $\omega$'s in (20) and (28) are independent on
the choice of $\Sigma$. It is important to mention that such
worldsheet conserved currents that play the role of integral
kernels for the symplectic structures on $Z$ in the present
approach, can be considered, in a more ordinary sense, as
Noetherian currents for the topological invariants considered,
which will allow us in this sense to obtain conserved currents
associated with any continuous symmetries of the background. In
the context of brane dynamics in the current literature, these
conserved currents only have been considered in this conventional
sense \cite{9}.

Since a symplectic structure $\omega$ governs the transition
between the classical and quantum domains, and allows us to
consider also the aspects of symmetry of the theory, it may be
interesting  to study the possible contribution of the topological
terms on the Poincar\'{e} charges, Poincar\'{e} algebra, and the
relevant commutation relations of the theory. In this sense,
because of the presence of the topological terms, the quantum
version of string theory obtained from the unmodified classical
dynamics, may be radically different to that obtained  from the
global description of the phase space given in terms of $\omega$.
All these questions will be the subject of forthcoming
communications. \\[2em]

\begin{center}
{\uno ACKNOWLEDGMENTS}
\end{center}
\vspace{1em}

The author acknowledges the financial support by the Sistema
Nacional de Investigadores (M\'exico), and wishes to thank A.
Escalante for conversations. \\[2em]

\end{document}